\documentclass[aps,prl,superscriptaddress,twocolumn,floatfix,showpacs,amsmath,amssymb]{revtex4}

\usepackage{graphicx}
\usepackage{amsmath}
\usepackage{amssymb}
\usepackage{color}
\usepackage{dcolumn}
\usepackage{epsfig}
\usepackage{bm}
\usepackage[urlcolor=blue]{hyperref}
\hypersetup{backref, colorlinks=true, linkcolor=blue, citecolor=blue}

\begin{document}
\title{Superconductivity in Efficient Thermoelectric Cu$_3$Sb$_{0.98}$Al$_{0.02}$Se$_4$}

\author{Xiao-Miao Zhao}
\affiliation{Center for High Pressure Science and Technology Advanced Research, Shanghai 201203, China}
\affiliation{Department of Physics, South China University of Technology, Guangzhou 510640, China}

\author{Yong-Hui Zhou}
\affiliation{Key Laboratory of Materials Physics, Institute of Solid State Physics, Chinese Academy of Science, Hefei 230031, China}
\affiliation{Center for High Pressure Science and Technology Advanced Research, Shanghai 201203, China}
\affiliation{Geophysical Laboratory, Carnegie Institution of Washington, Washington, DC 20015, U.S.A.}

\author{Qiao-Wei Huang}
\affiliation{Center for High Pressure Science and Technology Advanced Research, Shanghai 201203, China}

\author{Viktor V. Struzhkin}
\affiliation{Geophysical Laboratory, Carnegie Institution of Washington, Washington, DC 20015, U.S.A.}

\author{Ho-Kwang Mao}
\affiliation{Geophysical Laboratory, Carnegie Institution of Washington, Washington, DC 20015, U.S.A.}
\affiliation{Center for High Pressure Science and Technology Advanced Research, Shanghai 201203, China}

\author{Alexander G. Gavriliuk}
\affiliation{Institute of Crystallography, Russian Academy of Sciences, Leninsky pr. 59, Moscow 119333, Russia}
\affiliation{Institute for Nuclear Research, Russian Academy of Sciences, Troitsk, Moscow 142190, Russia}

\author{Xiao-Ying Qin}
\affiliation{Key Laboratory of Materials Physics, Institute of Solid State Physics, Chinese Academy of Science, Hefei 230031, China}

\author{Di Li}
\affiliation{Key Laboratory of Materials Physics, Institute of Solid State Physics, Chinese Academy of Science, Hefei 230031, China}

\author{Xi-Yu Li}
\affiliation{Key Laboratory of Materials Physics, Institute of Solid State Physics, Chinese Academy of Science, Hefei 230031, China}

\author{Yuan-Yue Li}
\affiliation{Key Laboratory of Materials Physics, Institute of Solid State Physics, Chinese Academy of Science, Hefei 230031, China}

\author{Xiao-Jia Chen}
\email{xjchen@hpstar.ac.cn}
\affiliation{Center for High Pressure Science and Technology Advanced Research, Shanghai 201203, China}
\affiliation{Key Laboratory of Materials Physics, Institute of Solid State Physics, Chinese Academy of Science, Hefei 230031, China}
\affiliation{Geophysical Laboratory, Carnegie Institution of Washington, Washington, DC 20015, U.S.A.}

\date{\today}

\begin{abstract}
Both superconductivity and thermoelectricity offer promising prospects for daily energy efficiency applications. The advancements of thermoelectric materials have led to the huge improvement of the thermoelectric figure of merit in the past decade. By applying pressure on a highly efficient thermoelectric material Cu$_3$Sb$_{0.98}$Al$_{0.02}$Se$_4$, we achieve dome-shape superconductivity developing at around 8.5 GPa but having a maximum critical temperature of 3.2 K at pressure of 12.7 GPa. The novel superconductor is realized through the first-order structural transformation from its initial phase to an orthorhombic one. The superconducting phase is determined in the ultimate formation of the Cu-Al-Sb-Se alloy. 
\end{abstract}

\pacs{73.63.Bd, 64.70.Nd, 74.70.Ad, 74.62.Fj}

\maketitle

Both superconductivity and thermoelectricity are two important phenomena for energy efficiency applications. The discovery of superconductivity at remarkably high temperatures in the layered copper oxides \cite{bedn} has spurred many subsequent discoveries of novel exotic superconductors such as the two-band superconductor MgB$_{2}$ with a critical transition temperature $T_{c}$ as high as 40 K \cite{naga}, exotic superconductivity in the heavy fermion superconductor PuCoGa$_{5}$ with a $T_{c}$ of 18.5 K \cite{sarr} (which is an order of magnitude higher than previously reported for this type of superconductor), and high-$T_{c}$ superconductivity in iron pnictides and chalcogenides \cite{kami} in which many structural, magnetic, nematic  orders coexist and compete with superconductivity. In addition, superconductivity has been discovered in carbon compounds like boron-doped diamond (11 K) \cite{ekim}, fullerides (33 K in Cs$_{x}$Rb$_{y}$C$_{60}$) \cite{tani}, and borocarbides (up to 16.5 K with metastable phases up to 23 K) \cite{cava} as well as in polycyclic aromatic hydrocarbons with a relatively high $T_{c}$ of 37 K \cite{ref7,ref10}. Pressure was found to play an important role in inducting superconductivity and enhancing $T_{c}$ in these materials \cite{jero,gani,chen2,lsun,saka,lgao}. For examples, the first organic superconductor was discovered in charge-transfer salts under pressure \cite{jero}. The application of pressure has also driven Cs$_{3}$C$_{60}$ from insulator to superconductor, with the highest \emph{T$_c$} of 38 K in fullerides \cite{gani}. The record high $T_{c}$s of 29 K in element superconductors \cite{saka} and 164 K in copper oxides \cite{lgao} were achieved at high pressures. The recent breakthrough in discovering superconductivity at 190 K in H-S system \cite{droz} further highlights the role of pressure. All these discoveries serve as springboards for the search for new superconducting materials. 

Thermoelectricity is about converting heat into electricity and vice-versa using the Seebeck and the Peltier effect, respectively. The efficiency of a thermoelectric material is characterized by the dimensionless figure of merit $zT$ which is determined by the electrical conductivity, Seebeck coefficient, and thermal conductivity at given temperature. The recent achievement of high $zT$ materials benefited from the technique developments such as disorder within the unit cell \cite{sale}, supplerlattices \cite{venk}, complex unit cells \cite{chun}, nanostructures \cite{12,11}, distortion of the electronic density of states \cite{here}, and ultralow thermal conductivity materials \cite{zhao}. The primary contribution to the $zT$ as high as the record 2.6 \cite{zhao} comes from the low thermal conductivity by scattering phonons. The $zT$ value of commercial material has been limited to about unit in all temperature ranges over early a half century \cite{snyd}. Cu$_3$SbSe$_4$, a narrow bandgap semiconductor with a bandgap of 0.1-0.4 eV \cite{1}, is being examined as an efficient thermoelectric material due to the following facts. First, it is Pb or Te-free thermoelectric material with less toxic and much easier to handle compared with other thermoelectric materials. Second, elemental substitution in Cu$_3$SbSe$_4$ has proved to be the most effective in raising $zT$, which increases from 0.7 (Cu$_3$Bi$_{0.02}$Sb$_{0.98}$Se$_4$ at 600 K) to 1.05 (Cu$_3$Sn$_{0.02}$Sb$_{0.98}$Se$_4$ at 690 K) \cite{7,10}. Third, nanostructured Cu$_3$SbSe$_4$ \cite{7} possesses the lower thermal conductivity and thus has higher $zT$, three times as large as that of the bulk material. Finally, this compound has high electrical conductivity, which contributes to its $zT$ as high as 1.2 at 550 K \cite{9}.

In this Letter, we show that pressure manipulates electronic behavior of an aluminium-doped Cu$_3$SbSe$_4$, a bulk material with high thermoelectric efficiency, and drives it to be a superconductor. The electrical transport, structural, and vibrational properties of this material are investigated by the combination of resistivity, synchrotron X-ray diffraction (XRD), and Raman scattering measurements. We demonstrate that the pressure-induced superconductivity is accompanied by the first-order structural transition. We thus obtain a novel superconductor from a dense high-$zT$ thermoelectric material.

The powder mixture from the very pure elements with the weighted according to the formula of Cu$_{3}$Sb$_{0.98}$Al$_{0.02}$Se$_{4}$ was loaded into quartz ampoules pumped under vacuum of 10$^{-2}$ Pa. The samples were slowly heated by 20 $^{o}$C/h for 45 hours up to 900 $^{o}$C and then held for 12 hours followed by the cooling to 500 $^{o}$C (1 $^{o}$C/min) and the quench in water at room temperature. The samples were then annealed at 300 $^{o}$C for 48 hours to promote homogeneity. The resulted ingots were pulverized into powders in an agate mortar. The bulk samples were obtained by spark plasma sintering for 5 min at temperature of 673 K and pressure of 50 MPa. The single-phase structure was confirmed from the XRD measurements. The fractographs were observed by field emission scanning electron microscopy. The typical grain size of the samples is $\sim$20 $\mu$m. Al substitution for Sb shrinks the host lattice and 2\% substitution shifts $zT$ increase by 30\% over wide temperatures up to 600 $^{o}$C. 

High-pressure electrical resistance measurements were performed by means of standard four-probe method in a miniature nonmagnetic diamond anvil cell \cite{alex}. A thin BN layer acted as an electric insulator between the electrodes leads and the gasket. Diamond anvil cells with T301 stainless steel gasket were used with the anvils in 300 $\mu$m culet for both XRD and Raman spectroscopy measurements. The XRD experiments at high pressures with synchrotron radiation were conducted at the Beijing Synchrotron Radiation Facility (BSRF) with a wavelength 0.6199 {\AA} at room temperature. Silicone oil was loaded as pressure transmitting medium to maintain quasi-hydrostatic pressure environment in two runs of XRD experiments. Pressure was measured by combining the ruby fluorescent method \cite{18} and the equation of states of Au \cite{19}. The sample-to-detector distance and experimental parameters were calibrated with standard CeO$_2$ powder diffraction. The two-dimensional diffraction images were converted to 2$\theta$ versus intensity data plots using the FIT2D software. The crystal structures were refined using GSAS package \cite{20}. The Raman spectra were measured in backscattering geometry with visible laser wavelength of 633 nm.

\begin{figure}[tbp]
\includegraphics[width=\columnwidth]{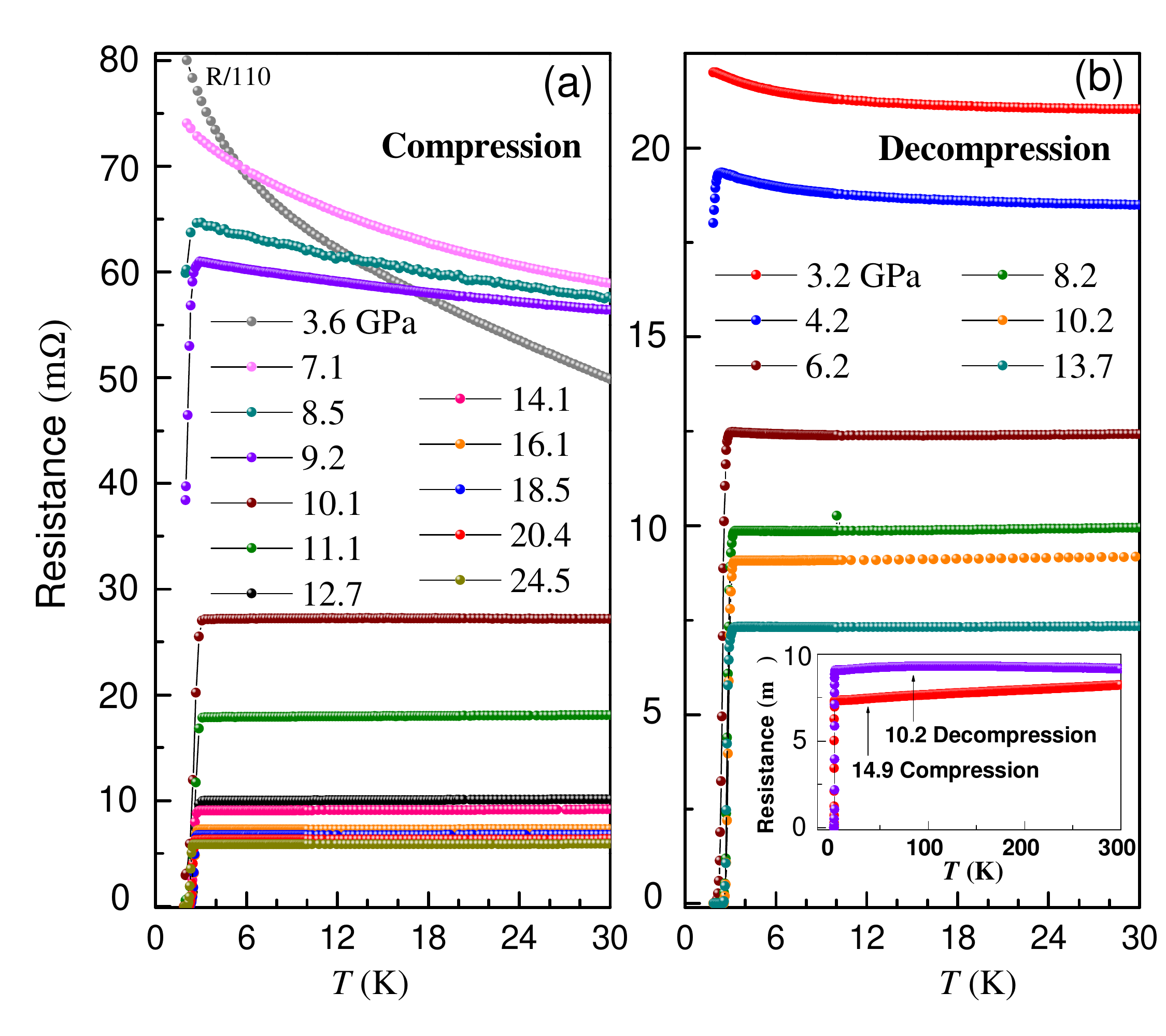}
\vspace{-0.5cm}
\label{raman} \caption{(Color online) Electrical resistance of Cu$_3$Sb$_{0.98}$Al$_{0.02}$Se$_4$ as a function of temperature at various pressures up to 24.5 GPa in both the compression (a) and decompression (b) runs. Inset shows the whole measured temperature range for the electrical resistance at 10.2 GPa (decompression) and 16.1 GPa (compression).}
\vspace{-0.5cm}
\end{figure}

We have measured the electrical resistance of Cu$_{3}$Sb$_{0.98}$Al$_{0.02}$Se$_4$ at high pressures and low temperatures. The results are shown in Fig. 1 for both the compression and decompression runs. We observed that the electrical resistance exhibits a semiconducting behavior in the low pressure region. For pressures above 8.5 GPa, a clear superconducting transition emerges with an onset \emph{T$_c$} at 2.3 K [Fig. 1(a)]. At the beginning, pressure-induced fraction of the the superconducting phase is not large enough for exhibiting zero resistance at 8.5 and 9.2 GPa. However, a sharp transition with a narrow width of around 0.2 K was soon reached upon further compression. The zero resistance is clearly observed, evidencing superconductivity in this material. These features indicate the good homogeneity of the superconducting phase. The value of \emph{T$_c$} exhibits a pronounced rise as pressure increases from 8.5 GPa, reaches a maximum of 3.2 K at 12.7 GPa, and then decreases gradually with the increase of  pressure. What is truly striking is that the resistance exhibits a semiconducting behavior over the whole temperature range at low pressures and keeps this behavior in the normal state when the material becomes superconductive below 10.1 GPa. This result follows that the material turns from a semiconductor to a semiconducting superconductor with the application of pressure. As pressure is further increased the resistance exhibits a nearly linear metallic decrease, contributing the metallic normal state. For the decompression run, it is rather unexpected to find that the superconducting phase can sustain to 4.2 GPa [Fig. 1(b)], which is much lower than the critical pressure of emergence of superconductivity in the compression run. An obvious hysteresis is thus obtained from the resistive measurements. The inset in Fig. 1(b) shows an expanded graph of resistance-temperature from 2 to 300 K for both the compression and decompression runs. 

\begin{figure}[tbp]
\includegraphics[width=\columnwidth]{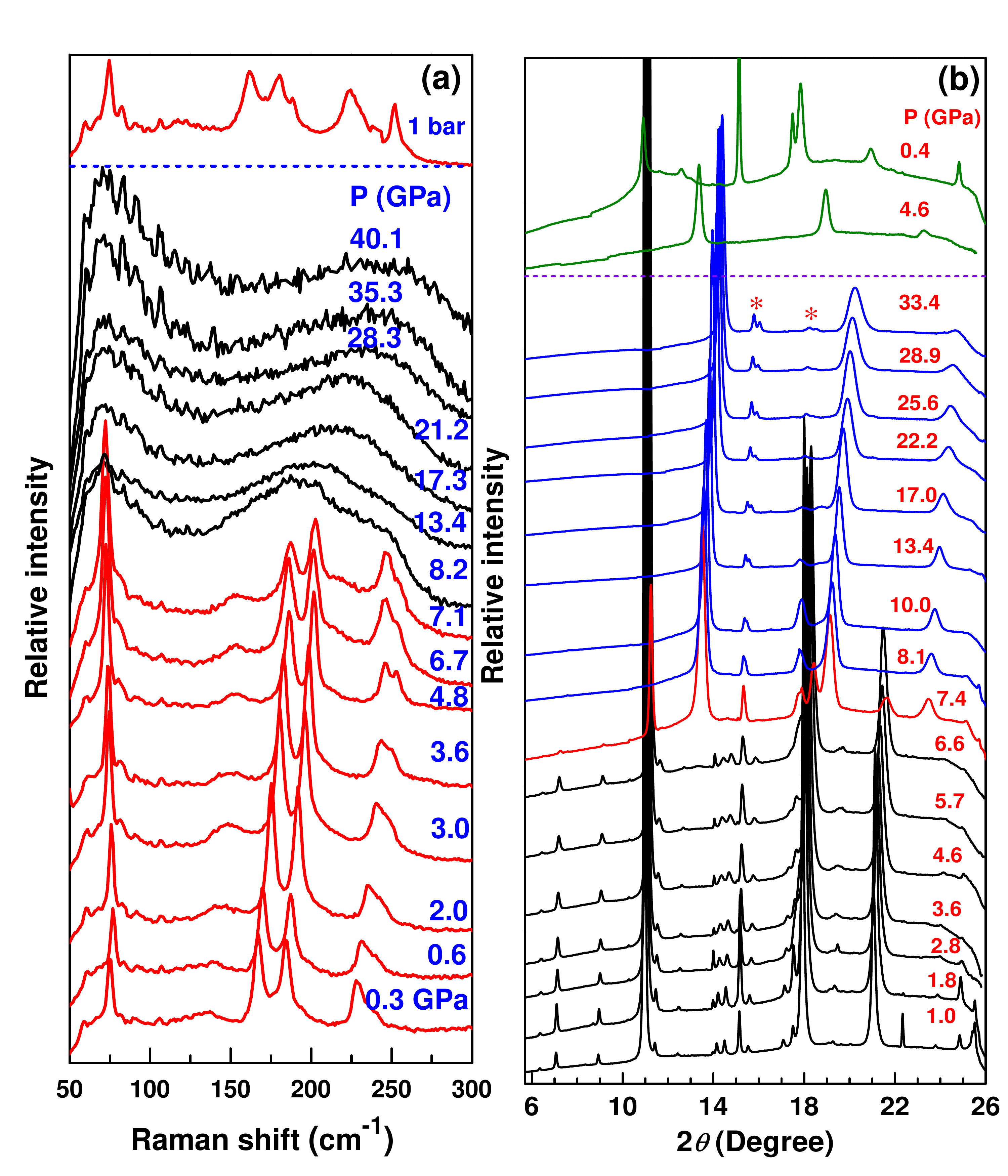}
\vspace{-0.5cm}
\label{modes} \caption{(Color online) Representative Raman spectra (a) and synchrotron x-ray powder diffraction patterns (b) of Cu$_{3}$Sb$_{0.98}$Al$_{0.02}$Se$_4$ at room temperature and pressures up to 40 GPa in both the compression and decompression runs. The diffraction signal of Au are marked with asterisks.}
\vspace{-0.5cm}
\end{figure}

We have performed high-pressure Raman scattering and synchrotron XRD measurements on Cu$_3$Sb$_{0.98}$Al$_{0.02}$Se$_4$ up to 40.1 GPa. The Raman spectra of this material at different pressures are shown in Fig. 2(a). When the pressure is increased to 8.2 GPa, there is an abrupt change of the Raman spectra, providing the evidence of phase transition. This observation provides an indication that phase transition may play a key role for superconductivity.  The clear and separate Raman modes disappear and a big bump emerges. It is worth noting that the big bump, which is corresponding to the overlap of multiple peaks, shifts gradually toward higher frequency with increasing pressure. Upon decompression, we observed that the sample recovers to its initial phase at ambient conditions.

Figure 2(b) presents the XRD patterns of Cu$_3$Sb$_{0.98}$Al$_{0.02}$Se$_4$ for applied pressure ranging from 1 to 33.4 GPa. All the Bragg peaks shift to larger angles, showing the shrinkage of the lattice constant. Upon compression, as we expected, there are several changes in the XRD patterns in the number, intensity, and shape of the peaks, suggesting that an obvious structural transition takes place above 8.1 GPa. Upon further compression, the XRD patterns have no change up to 33.4 GPa. The diffraction data in low-pressure region can be fitted using the space group $\emph{I}$$\bar{4}2/$$\emph{m}$, which is consistent with the crystal structure of Cu$_3$SbSe$_4$ \cite{22}. The critical pressure of structural transition from XRD data shows an excellent agreement with the Raman spectra. This result provides an opportunity to better understand the underling mechanism of superconductivity in Cu$_3$Sb$_{0.98}$Al$_{0.02}$Se$_4$. Upon decompression, the crystal structure remains the high-pressure phase at 4.6 GPa and completely returns to the low-pressure phase at 0.4 GPa, suggesting a tiny but noticeable trace of hysteresis. Peaks marked with asterisks are the diffraction signal of Au in Fig. 2(b). In order to accurately determine the crystal structure of high-pressure phase, we repeated the XRD measurements of Cu$_3$Sb$_{0.98}$Al$_{0.02}$Se$_4$ without Au, and the diffraction patterns were the same as the first one. Both Raman scattering and XRD data under pressure provide consistent evidence for the pressure-induced structural transition and its association with superconductivity. 

\begin{figure}[tbp]
\includegraphics[width=\columnwidth]{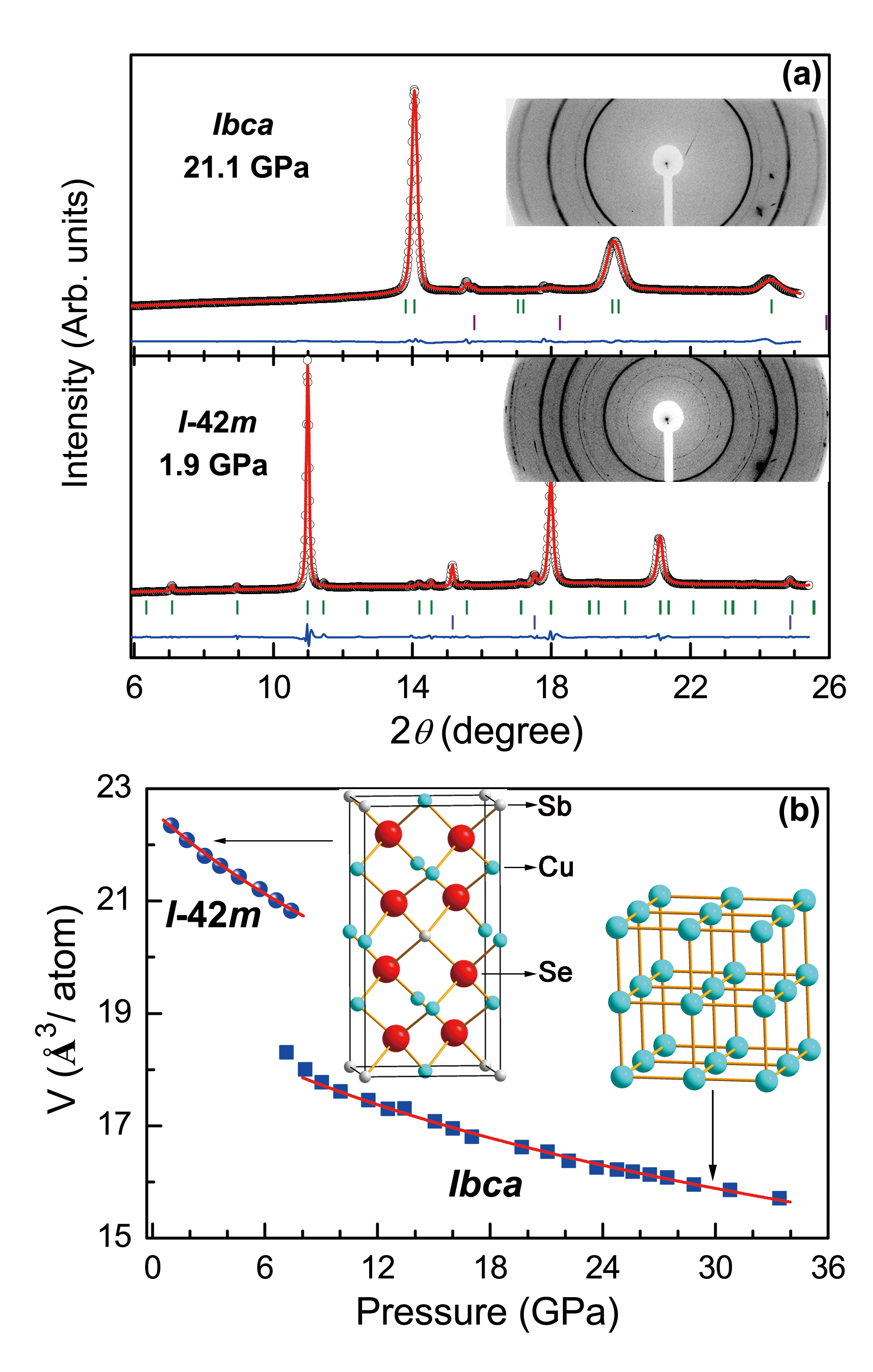}
\caption{(Color online) (a) XRD patterns of Cu$_3$Sb$_{0.98}$Al$_{0.02}$Se$_4$ at pressures of 1.9 and 21.1 GPa for the corresponding space groups. The open circles represent the measured intensities and the red lines the results of profile refinements by the best LeBail fit with each space group. The positions of the Bragg reflections are marked by vertical lines and the difference profiles are shown at the bottoms (blue lines). The \emph{R} values are \emph{R}$_{\emph{p}}$ = 1.7\%, \emph{R}$_{\emph{wp}}$=2.7\% for the fitting at 21.1 GPa, \emph{R}$_{\emph{p}}$=1.0\%, \emph{R}$_{\emph{wp}}$=1.5\% at 1.9~GPa.
(b) Pressure dependence of the volume per atom of Cu$_3$Sb$_{0.98}$Al$_{0.02}$Se$_4$ in both low-pressure phase and high-pressure phase. Solid lines correspond to the results of a least-squares fit using equation of states. The insets illustrate the atomic arrangement of the low-pressure and high-pressure structures.}
\vspace{-0.5cm}
\end{figure}

Next we present the detailed refinement of the diffraction data of high-pressure phase. Cu$_3$SbSe$_4$ is considered as a four-fold ternary derivative of ZnSe. After structural transition, there are only three diffraction peaks when 2$\theta$ is below 26$^{o}$ for Cu$_3$Sb$_{0.98}$Al$_{0.02}$Se$_4$. The XRD pattern at 8.1 GPa was indexed with a structure similar to the cubic structure, but it is evidenced not a standard strictly cubic structure. Three lattice parameters are very close but not completely equal and present different compression coefficients with increasing pressure. This is further demonstrated by the gradual broaden diffraction peaks formed from the overlap of multiple diffraction peaks. The further analysis indicates that the structural arrangement of the high-pressure phase is similar to that of $\emph{Fm}$3$\emph{c}$ or $\emph{F}$$\bar{4}$3$\emph{c}$, except for the very small difference of lattice parameters. Through indexing and refining these data, we determined the high-pressure phase belonging to a space group $\emph{Ibca}$, which can be recognized as a deformation of the cubic structure to an orthorhombic unit cell. The theorists predicted that the similar structure material of ZnSe and ZnTe transform into orthorhombic phase finally under pressure with the minimum total energy, providing further evidence to support our result \cite{221}.

The fitted results of both the low-pressure phase with $\emph{I}$$\bar{4}2/$$\emph{m}$ space group at 1.9 GPa and the high-pressure phase with $\emph{Ibca}$ space group at 21.1 GPa are shown in Fig. 3(a), respectively. We found that Cu$_3$Sb$_{0.98}$Al$_{0.02}$Se$_4$ forms an Al-Cu-Sb-Se substitutional alloy after the phase transition is complete, similar to the behavior of Bi$_2$Te$_3$ at high pressures \cite{23}. The position at (0,0,0) of space group $\emph{Ibca}$ was occupied by the Al-Cu-Sb-Se alloy, as obtained from powder X-ray data. The pressure dependence of the unit-cell volume $V$ is usually described by the three order Birch-Murnaghan equation of state \cite{24}. Fitting our data points yields the bulk modulus \emph{K}$_0$ = 73.73$\pm$0.01 GPa and its pressure derivative \emph{K}$_0$$^\prime$ = 5.2$\pm$0.01 for low-pressure phase, and the ambient-pressure volume $V_0$ = 22.62$\pm$0.01 {\AA}$^3$. The fit to the data within the high pressure phase results in values of \emph{K}$_0$ = 89.30$\pm$0.24 GPa, \emph{K}$_0$$^\prime$ =6.18$\pm$0.02, and $V_0$ = 19.19$\pm$0.01 {\AA}$^3$. It is important to note that the value of \emph{K}$_0$ obtained from $\emph{I}$$\bar{4}2/$$\emph{m}$ is smaller than the obtained \emph{K}$_0$ in $\emph{Ibca}$, suggesting the harder bonds at high-pressure phase. The structural transition leads to the contraction of the volume per atom. It is established \cite{25} that the volume is decreased by 21\% for Zinc sulphide and the 28\% for Zinc selenide during the structural transition, respectively, which is larger than the collapse in volume about 15.1 \% from our data for Cu$_3$Sb$_{0.98}$Al$_{0.02}$Se$_4$.

\begin{figure}[tbp]
\includegraphics[width=\columnwidth]{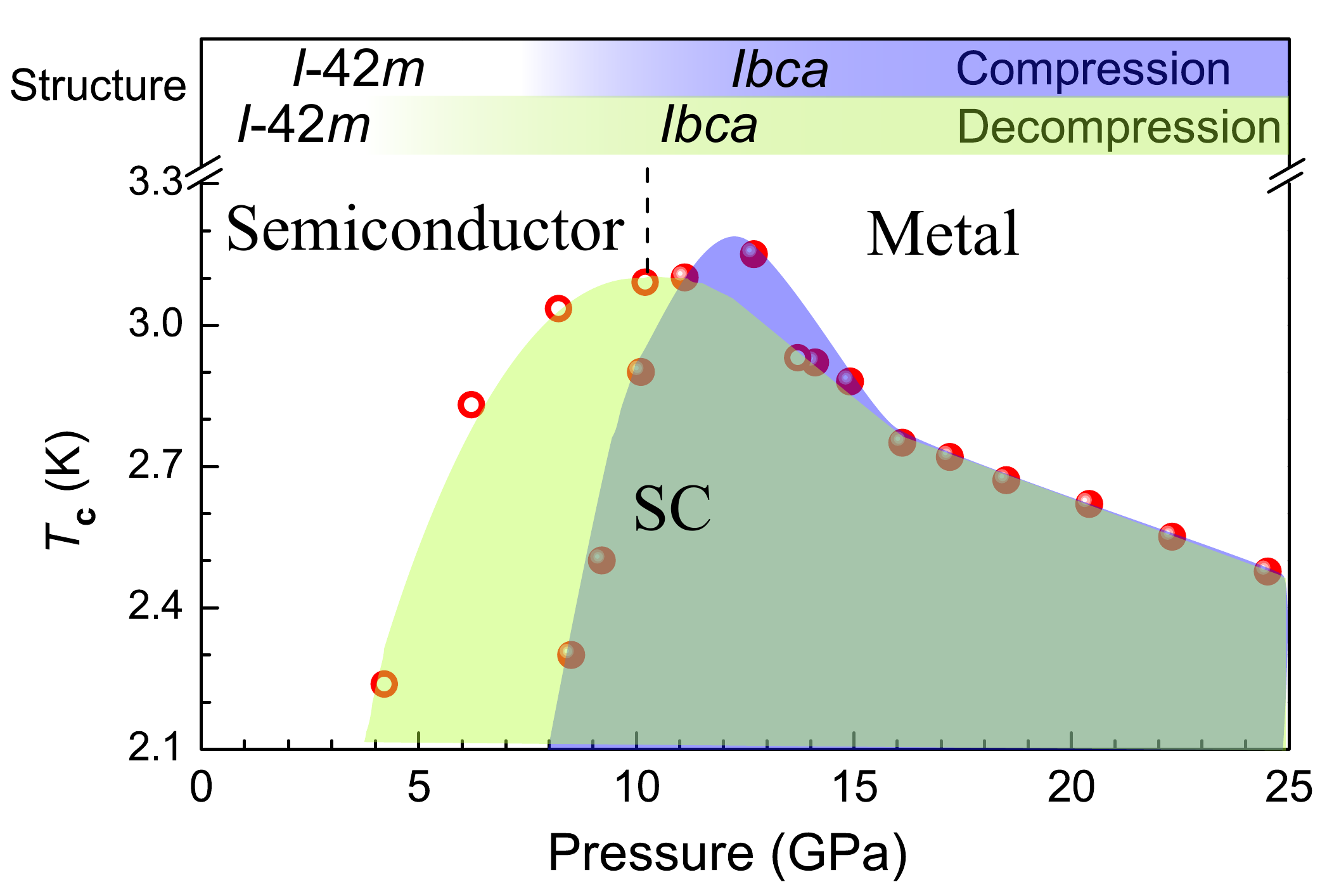}
\vspace{-0.6cm}
\caption{(Color online) Phase diagram of Cu$_3$Sb$_{0.98}$Al$_{0.02}$Se$_4$ showing the evolution of  \emph{T$_c$} and crystal structure with pressure in both the compression and decompression runs.}
\vspace{-0.4cm}
\end{figure}

We summarized both the variation of \emph{T$_c$} and structural evolution as a function of pressure up to 25 GPa in Fig. 5. The phase diagram clearly reveals the close correlation between the structural transition and the appearance of superconductivity. Upon compression, an interesting structural transition at 8.1 GPa from $\emph{I}$$\bar{4}2/$$\emph{m}$ symmetry to $\emph{Ibca}$ symmetry was observed, below which there is no signature of superconducting transition. Note that \emph{T$_c$} is enhanced steeply as the pressure increases from 8 to 13 GPa, then it decreases with applied pressure after reaching the maximum value of 3.2 K. After that, there is a monotonous reduction of \emph{T$_c$} initially at a rate of 0.12 K GPa$^{-1}$ as pressure is increased, however, the rate of \emph{T$_c$}-decrease is slowed down with pressure above 16 GPa. Upon decompression, the superconducting phase disappears below 4.2 GPa with the effect of hysteresis, which is the same as the behavior of structure. 

Over the whole pressure range studied, superconductivity correlates well with crystal structure on both the compression and decompression runs. Below 10 GPa, an increase of resistivity with decreasing temperature is observed either in the measured temperature range or in the normal state. This indicates that Cu$_3$Sb$_{0.98}$Al$_{0.02}$Se$_4$ undergoes the change from semiconductor to superconductor at around 8.5 GPa. Superconductivity is realized in this material from semiconductor rather than metal. Such a behavior has been observed in many narrow bandgap semiconductors \cite{26,27,28,29}. The observed superconductivity in Cu$_3$Sb$_{0.98}$Al$_{0.02}$Se$_4$ may have the similar origin. For semiconducting superconductors, $T_{c}$ is mainly controlled by carrier concentration \cite{30}. Superconductivity in these materials arises primarily from the attractive interaction resulting from the exchange of intravalley and intervalley phonons, which can be larger than the repulsive Coulomb interaction. 

In summary, we have reported a finding of pressure-induced superconductivity in the thermoelectric material Cu$_3$Sb$_{0.98}$Al$_{0.02}$Se$_4$. The detailed Raman scattering and synchrotron XRD measurements revealed the close relationship between the structural transition and the emergence of superconductivity. The superconducting phase was determined to be a alloy with an orthorhombic structure. These results suggest that thermoelectric materials offer a promising platform for the exploration of novel superconductors. Recent first-principles calculations \cite{6} predicted that this class of ternary famatinite is a candidate for topological insulators. Investigating the interplay of the topological order, thermoelectricity, and superconductivity of these materials at high pressures would be an interesting topic.

\begin{acknowledgments}
This work was supported at SCUT by the Cultivation Fund of the Key Scientific and Technical Innovation Project Ministry of Education of China (Project No. 708070) and at Carnegie by EFree, an Energy Frontier Research Center funded by the U.S. Department of Energy (DOE), Office of Science, Office of Basic Energy Sciences (BES) under Award Number DE-SG0001057. The resistance measurements were supported by the DOE under Grant No. DE-FG02-02ER45955. Use of the BSRF for XRD measurements was supported by the Chinese Academy of Sciences. The sample design and growth were supported by the Natural Science Foundation of China (Nos. 11174292 and 11374306). A.G.G. acknowledges the support from Russian Foundation for Basic Research grant 14-02-00483-a, from RAS programs ``Strongly correlated electron systems'', ``Elementary particle physics, fundamental nuclear physics and nuclear technologies''.
\end{acknowledgments}

\end{document}